\documentclass{ws-procs9x6}

\begin{document}

\title{Exotic superfluids: Breached Pairing, Mixed Phases and Stability
\footnote{\uppercase{T}his work was performed in collaboration
with \uppercase{M.~F}orbes, \uppercase{W.~V.~L}iu and \uppercase{F.~W}ilczek.}}

\author{E. GUBANKOVA}

\address{Center for Theoretical Physics, Department of Physics,\\
MIT, Cambridge, Massachusetts 02139}

\maketitle

\abstracts{We review properties of gapless states. We construct a model
where a stable breached pair (gapless) state is realized.}

\section{Introduction}

Motivated by recent experiments in cold atoms \cite{Nature} and by questions in QCD
at high densities \cite{Review}, we consider here superfluid fermion systems, in particular
exotic superfluids. As suggested by Liu and Wilczek, exotic phases of matter in superfluid fermion systems
involve coexistence of normal Fermi liquid and superfluid components \cite{LiuWilczek}.
Superfluid properties are described by a nonzero condensate,
$\Delta=\langle T(\psi^{\dagger}(x)\psi^{\dagger}(x^{\prime}))\rangle \neq 0$,
being the order parameter, while a single quasiparticle dispersion crosses the momentum axis
(free Fermi surface) leading to a gapless mode; thus this phase is sometimes called a
``gapless'' superconductive phase. Due to nonzero Fermi condensate the ground state
is a superfluid in a classical sense (\textit{i.e.} it has zero viscosity).
On a microscopic scale, one can envision a momentum separation in exotic superfluids.
For species with noticeably different Fermi momenta, Cooper pairing takes place around
the Fermi surfaces, but there is no pairing in the momentum region between surfaces (the \textit{breach}).
Thus in this work we use the term ``breached pair'' (BP) superfluidity.

To obtain exotic superfluids we consider pairing between two different
fermion species whose Fermi surfaces do not match. This possibility arises in several
situations: (1) Spin-up spin-down electrons in an ordinary superconductor placed in a uniform
magnetic field undergo Zeeman splitting, leading to a mismatch in Fermi momentum.
As found by Sarma in the 1960's \cite{Sarma}, an exotic superconducting ground state should arise
at large momentum
mismatch when $\mu_B H>\Delta$. Before that, however, the first-order phase transition from
the superconducting state to the normal state takes place at $\mu_B H=\Delta/\sqrt{2}$.
Placing a superconductor in a spacially varying magnetic field
or adding paramagnetic impurities
with a strong spin-flip electron-impurity scattering amplitude stabilizes the gapless
superconductor \cite{AbrikosovGorkovDzyaloshinski}.
(2) Recent experiments
in cold atomic fermion gases trapped in an optical lattice and operating near Feshbach
resonance deal with a mixture of two hyperfine spin components of alkali atoms. By changing
the scattering length one can go from the regime of Bose-Einstein condensation to
BCS superfluidity, which is of interest in this work.
Laser lattice involves counterpropagating laser beams, that together generate a standing light
wave leading to different AC Stark shifts for the spin-up and spin-down components. Using methods
of 'engineering' various lattice systems and by tuning effective masses one can produce
exotic phases \cite{LiuWilczekZoller}.
(3) In strongly interacting quark matter at high baryon densities and low temperatures,
different flavors of quarks pair and form color superconductors \cite{Review}. Here
a mismatch in Fermi momenta arises due to a nonzero strange quark mass, $m_s\neq 0$, and is
triggered by imposing a charge neutrality condition.
At intermediate density (2--3 nuclear densities), an
exotic state may arise \cite{AlfordKouvarisRajagopal} which links the CFL and nuclear matter phases.

Plan: First, we give a general analysis of gapless states. Then, we show how to realize
a stable BP superfluid state.

\section{Gapless state and its stability}

\begin{figure}[htb]
\centerline{\epsfxsize=4.1in\epsfbox{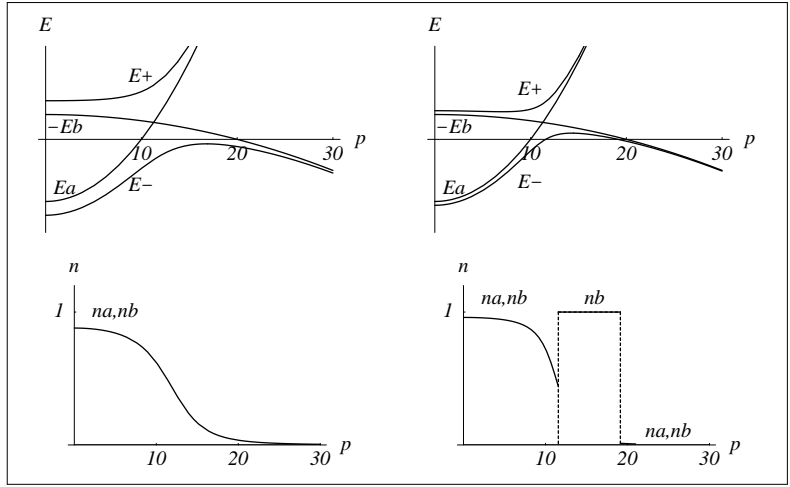}}
\caption{Dispersion relations and occupation numbers for the BCS (left)
and Sarma (right) states. \label{picture1}}
\end{figure}

We consider the mean-flield analysis of a model with two species of fermions A, and B
of differing masses $m_{A}<m_{B}$ and in the absence of interaction with different
Fermi momenta $p_{A}<p_{B}$. The Hamiltonian
is $H=\int d^{3}p/(2\pi)^3\left(\varepsilon_{p}^{A}\psi^{\dagger}_{Ap}\psi_{Ap}
+\varepsilon_{p}^{B}\psi^{\dagger}_{Bp}\psi_{Bp}\right)+H_{I}$
with attractive interaction $H_{I}=-g\int d^{3}p/(2\pi)^3 d^{3}q/(2\pi)^3
\psi^{\dagger}_{Ap}\psi^{\dagger}_{B-p}\psi_{Bq}\psi_{A-q}$ ($g>0$)
and $\varepsilon_{p}^{A}=p^2/2m_{A}-\mu_{A}$, $\varepsilon_{p}^{B}=p^2/2m_{B}-\mu_{B}$,
where $p_{A}=\sqrt{2m_{A}\mu_{A}}$, $p_{B}=\sqrt{2m_{B}\mu_{B}}$.
At the mean-field level, the condensate $\Delta=\int d^3p/(2\pi)^3
\langle \psi_{Bp}\psi_{A-p}\rangle$ is a c-number, which permits to diagonalize
the Hamiltonian. As a result, the quasiparticle excitations
are $E^{\pm}_{p}=\varepsilon_p^{-}\pm\sqrt{\varepsilon_p^{+\,2}+\Delta^2}$
with $\varepsilon_p^{\pm}=(\varepsilon_p^{A}\pm\varepsilon_p^{B})/2$;
they contain mixture of A-particle and B-hole excitations.
We minimize the thermodynamic potential
$\Omega=H-\mu_A n_A -\mu_B n_B$, i.e. $\partial\Omega/\partial\Delta =0$,
to find the gap parameter $\Delta$. There are two non-trivial solutions.
First one with larger gap corresponds to a fully gapped BCS state, where $E^{\pm}_{p}$
have opposite signes for all momenta. Though $n_{A}\neq n_{B}$ at $g=0$, in the presence
of interaction particles redestribute so that the occupation numbers become equal
$\tilde{n}_{A}=\tilde{n}_{B}=n$ with
$n=1/2\left(1-\varepsilon_p^{+}/\sqrt{\varepsilon_p^{+\,2}+\Delta^2}\right)$;
the BCS condensation energy is the largest for
$\tilde{p}_{A}=\tilde{p}_{B}$. Second solution, obtained by Sarma in 60's,
has smaller gap. Interaction is not strong enough to pull both Fermi surfaces
together, leaving a breach region with single occupancy by B-particles.
Pairing and superfluidity takes place primarily around the smaller Fermi surface, while
there is normal component for momenta in the breach region
where $E^{\pm}_{p}$ have the same signs (separation in momentum).
Points where $E^{-}_{p}=0$, i.e.,
$p_{1,2}^2=(p_{A}^2+p_{B}^2)/2\pm 1/2\sqrt{(p_{B}^2-p_{A}^2)^2-16m_{A}m_{B}\Delta^2}$,
give the free Fermi surfaces
leading to gapless modes and to free fermion liquid component. Occupation
numbers are $\tilde{n}_{A}=\tilde{n}_{B}=n$ for $0\leq p\leq p_{1}$ and $p_{2}\leq p$,
and $\tilde{n}_{B}=1$ $\tilde{n}_{A}=0$ for $p_{1}\leq p\leq p_{2}$,
Figure~\ref{picture1}.

We consider nontrivial superfluid solutions of the gap equation in grand canonical
and canonical ensembles.

\begin{figure}[htb]
\centerline{\epsfxsize=4.1in\epsfbox{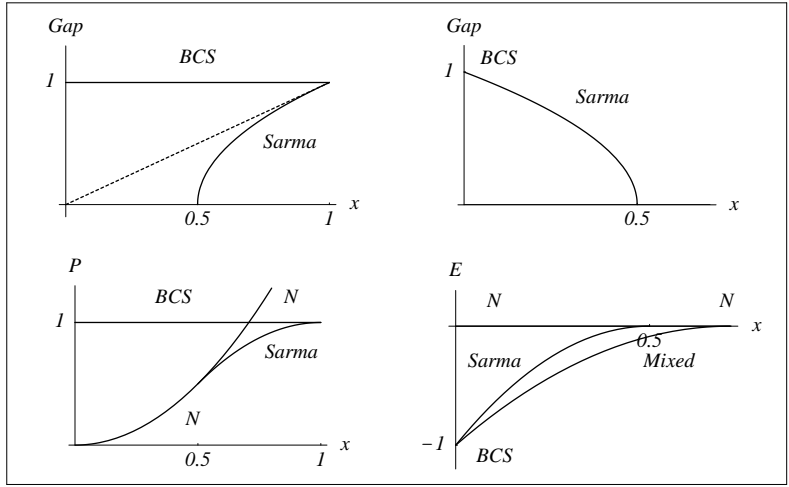}}
\caption{Solutions of the gap equation and corresponding pressures as functions
of the Fermi momenta mismatch at fixed chemical potentials (left), and
gap equation solutions and energies at fixed particle numbers (right). \label{picture2}}
\end{figure}

I. Grand canonical ensemble.

We fix the chemical potentials
$\mu_{A}, \mu_{B}$ and minimize the thermodynamic potential
$\Omega_{S}=H-\mu_{A}n_{A}-\mu_{B}n_{B}$ over all ground state
superconducting wave functions,
$min\langle \Psi_{S}|\Omega|\Psi_{S}\rangle$. We obtain,
apart from the trivial solution $\Delta=0$, two nontrivial solutions:
the fully gapped BCS and the gapless Sarma solution \cite{Sarma}. As a function of the Fermi
momenta mismatch $\delta p=(p_{B}^2-p_{A}^2)/4\sqrt{m_{A}m_{B}}$,
they are $\Delta_{BSC}=\Delta_0$ and
$\Delta_{Sarma}/\Delta_{0}=\sqrt{2x-1}$ with $x=\delta p/\Delta_{0}$, Figure~\ref{picture2}.
Since at a given $\delta p$
and $g$, $\Delta_{BCS}>\Delta_{Sarma}$, BCS state wins, i.e., system prefers
the BCS ground state over the Sarma state.
At fixed Fermi momenta $p_{A}, p_{B}$, the thermodynamic potential
as a function of the gap has two minima -- normal (N) and the BCS (absolute min) states,
and one maximum -- the Sarma state. Hence the Sarma state is metastable.

We consider the pressure versus the Fermi momentum mismatch for states which are
solutions of the gap equation, where normalized pressure is defined through
the condensation free energy as
$P_{S}=-(\langle\Omega_{S}\rangle-\langle\Omega_{0}\rangle)
/|(\langle\Omega_{BCS}\rangle-\langle\Omega_{0}\rangle)|$
where $\langle\Omega_{0}\rangle$ is the free
energy of the normal state at $\delta p=0$ and the BCS condensation energy is
$\langle\Omega_{BCS}\rangle-\langle\Omega_{0}\rangle = -N(0)\Delta_0^2/2$,
$\Delta_0$ is the BCS gap and $N(0)$ is the density of states at the Fermi surface.
In the leading order $\Delta\sim \delta p\ll p_{A}, p_{B}$
(i.e., when all quantities are written as expansions near the Fermi surface),
pressures are
$P_{BCS}=1$ for $0\leq x\leq 1$,
$P_{Sarma}=2x^2-(1-2x)^2$ for $1/2\leq x\leq 1$ and $P_{N}=2x^2$
for $x\geq 0$ with $x=\delta p/\Delta_0$, Figure~\ref{picture2}.
State with maximum pressure
(or minimum condensate energy) wins. For $p_{A}=p_{B}$
there is the BCS state. As we add B-particles, we create stress, and
pressure of the BCS state $P_{BCS}$ drops relatively to the pressure of the normal unpaired
state $P_{N}\sim \delta p^2$. When $P_{BCS}-P_{N}\leq 0$, there is a first order
phase transition at $\Delta\neq 0$ from the superconducting to the normal state.
The BCS superconductor is destroyed when win from condensation energy besomes less than loss
in energy needed to pull the Fermi surfaces together to create the BCS.
Sarma state is tangent to normal state at $\Delta=0$, and has always lower pressure
than normal state; hence Sarma state is unstable.

II. Canonical ensemble.

We fix the particle numbers $n_{A}, n_{B}$, allowing
the chemical potentials to change,
and minimize the Helmholtz energy over all possible superconducting ground states
with constraint of fixed $n_{i}$ $i=A, B$,
$min\langle \Psi_{S}|H|\Psi_{S}\rangle_{n_{i}-const}$ \cite{GubankovaLiuWilczek}.
At a single point when $n_{A}=n_{B}$
there is the BCS state with $\Delta_{BCS}=\Delta_{0}$.
At $n_{A}\neq n_{B}$, there is Sarma state with decreasing gap as
$\delta n=n_{B}-n_{A}$ increases,
$\Delta_{Sarma}/\Delta_{0}=\sqrt{1-2x}$ where $x=\delta p/\Delta_{0}\sim\delta n$.
The energy of Sarma state is lower than the energy of the normal state, $E_{Sarma}<E_{N}$;
thus Sarma state is stable.
Imposing neutrality condition a stable gapless superconducting state was obtained
in the QCD context by Shovkovy et. al. \cite{ShovkovyHuang}.

We came to different conclusions about stability of Sarma state
using grand canonical and canonical ensembles. This difference in stability
analysis can be resolved by considering mixed phase, which is a mixture in space of two
(or more) homogeneous states. Bedaque et. al. \cite{BedaqueCaldasRupak} suggested to consider a mixture
of the BCS and normal states, separated in $x$-space.
They found that
the energy of the mixed state is lower than the energy of Sarma state,
$E_{mixed}<E_{Sarma}$; thus Sarma state is unstable with respect to decay into a mixed state.
We confirm their findings.
We define the normalized condensation energy as
$E_{S}=(\langle H_{S}\rangle-\langle H_{N}\rangle)/|(\langle H_{BCS}\rangle -\langle H_{N}\rangle)|$
where $\langle H_{BCS}\rangle-\langle H_{N}\rangle=-N(0)\Delta_0^2/2$.
In the leading order $\Delta\sim \delta p\ll p_{A}, p_{B}$,
condensation energies are
$E_{BCS}=-1$ for $x=0$, $E_{Sarma}=-(1-2x)^2$ for $0\leq x\leq 1/2$,
$E_{mixed}=-(1-\sqrt{2}x)^2$ and $E_{N}=0$, $x=\delta p/\Delta_0$
and $\Delta_0$ is the BCS gap, Figure~\ref{picture2}.

\begin{figure}[htb]
\centerline{\epsfxsize=4.1in\epsfbox{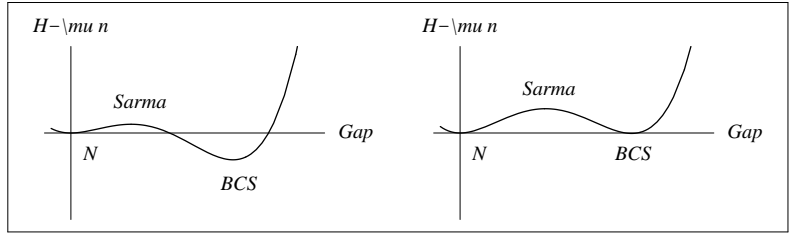}}
\caption{Thermodynamic potential as a function of the gap,
and positions of each state. Sarma state decays into a mixed
BCS and Normal states depicted on the right panel.
\label{picture3}}
\end{figure}

Allowing mixed states in our ansatz of trial ground state wave functions $\Psi_{S}$,
metastable Sarma state decays (rolls down) into a mixture of
the BCS and normal states, Figure~\ref{picture3}.
(In a mixed state, pressures and chemical potentials of composite states are equal,
hence there are two equal minima $\Omega_{BCS}=\Omega_{N}$).
Generally, it is difficult to include mixed states in variational ansatz
in the grand canonical ensemble.

We conclude, that in grand canonical and canonical ensembles Sarma branch
of the gapless solutions is unstable.

In \cite{ForbesGubankovaLiuWilczek}, we show that conclusion about stability
of a state is the same in any ensemble used. In particular, there is one-to-one correspondence
between a state at fixed particle number(s) and the state that minimizes
the thermodynamic potential $\Omega$ in grand canonical ensemble. Thus,
there is always a stable state in the grand canonical ensemble that
satsfies the constraint. Imposing constraint (over particle numbers)
cannot stabilize the system. Practical guide is to look for a stable (gapless) solution
in the grand canonical ensemble.

\section{Breached paired superfluid state for a finite-range interaction}

Our goal is to construct a stable breached paired state which has coexisting superfluid
and gapless components.
We use an idea
that existance of the gapless modes depend on the momentum structure
of the gap $\Delta(p)$. There should be two distinct regions in momentum space:
first one where $\Delta_p$ is large enough to support the superfluid,
and second one where $\Delta_p$ is small enough that pairing does not appreciably
affect the normal free-fermion behavior. To garantee stability, the phase
must also have higher pressure than the normal state. We realize such BP states
in two examples \cite{ForbesGubankovaLiuWilczek}.

\begin{figure}[htb]
\centerline{\epsfxsize=4.1in\epsfbox{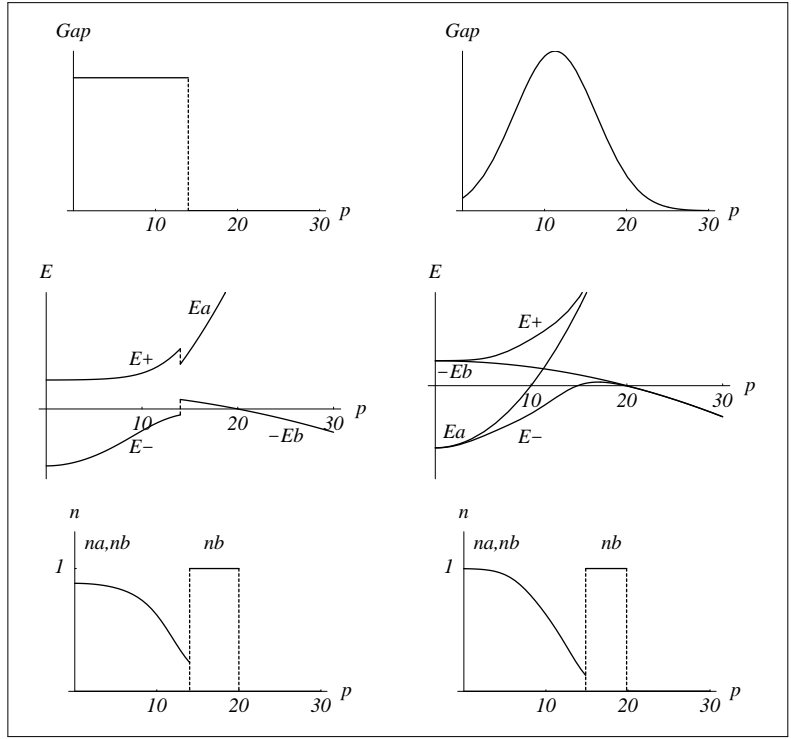}}
\caption{Gap, dispersions and occupation numbers for the cut-off (left)
and the two-body potential (right) interactions.
\label{picture4}}
\end{figure}

I. Cut-off interaction.

We impose a cut-off interaction such that
it supports the BCS-like pairing for $p<p_{\Lambda}$
and it allows free dispersion relations for $p>p_{\Lambda}$,
accomodating the excess of B-particles and leading to gapless modes.
We construct this state by minimizing the thermodynamic potential,
$min\langle\Psi_{S}|H-\mu_{A}n_{A}-\mu_{B}n_{B}|\Psi_{S}\rangle$,
where $H$ includes the cut-off interaction
$-g\int d^3p/(2\pi)^3 d^3q/(2\pi)^3f(p)f(q)
\psi^{\dagger}_{Ap}\psi^{\dagger}_{B-p}\psi_{B-q}\psi_{Aq}$
with $f(p)=1$ for $p>p_{\Lambda}$ and $f(p)=0$ for $p\leq p_{\Lambda}$.
The gap parameter, defined as
$\Delta=g\int d^3p/(2\pi)^3f(p)\langle \psi_{Bp}\psi_{A-p}\rangle$,
satisfies the gap equation
$\Delta=1/2g \int d^3q/(2\pi)^3\Delta f(q)/\sqrt{\varepsilon_p^{+\,2}+\Delta^2}$
where momentum integration is performed outside the breach region.
The occupation numbers $n_{A}, n_{B}$ show the evidance
that it is a breached paired state, Figure~\ref{picture4}.
This state is an absolute minimum
of the thermodynamic potential, hence we obtained
a stable BP state.

II. Spherically symmetric static two-body potential.

With attractive potential $V(x-x^{\prime})$,
interaction is $H_{I}=\int d^3p/(2\pi)^3 d^3q/(2\pi)^3 V(p-q)
\psi^{\dagger}_{Ap}\psi^{\dagger}_{B-p}\psi_{B-q}\psi_{Aq}$ and
the gap parameter acquires a momentum dependence,
$\Delta_p=\int d^3 q/(2\pi)^3V(p-q)\langle \psi_{Bq}\psi_{A-q}\rangle$.
The gap equation is written
$\Delta_p=1/2\int d^3q/(2\pi)^3v(p-q)\Delta_q/\sqrt{\varepsilon_q^{+\,2}+\Delta_q^2}$,
and quasiparticle dispersion relations are
$E_p^{\pm}=\varepsilon_p^{-}\pm\sqrt{\varepsilon_p^{+\,2}+\Delta_p^2}$.
We take a gaussian potential for numerical simulations.
Due to the BCS instability, $\Delta_p$ picks at the effective Fermi surface
given by the pole of the gap equation at $\Delta=0$, $\varepsilon_{p_{0}}^{+}=0$.
Therefore $\Delta_p$ supports the BCS-like pairing around $p_0$,
and allows free dispersion relations, and hence free Fermi surfaces,
outside the breached region, Figure~\ref{picture4}. It is, however, difficult to varify
that this state is an absolute minimum of the thermodynamic potential
since instead of a number, $\Delta$, we have a function $\Delta_p$
in the variational ansatz.

\begin{figure}[htb]
\centerline{\epsfxsize=4.1in\epsfbox{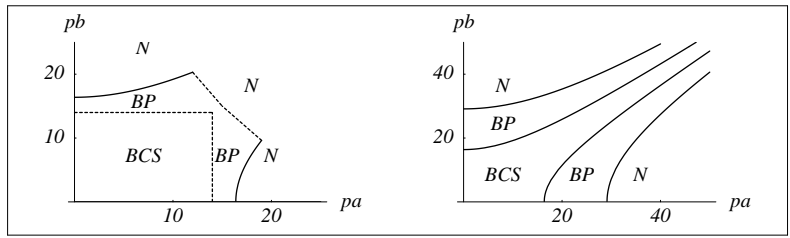}}
\caption{Phase diagram of possible homogeneous phases in coordinates of the Fermi momenta
$(p_A,p_B)$ for the cut-off (left) and two-body potential (right) interactions.
\label{picture5}}
\end{figure}

We performed minimization of the thermodynamic potential numerically
using different potentials. Generally, there is a central
strip of fully gapped BCS phase about $p_{A}=p_{B}$ with normal unpaired
phase outside. Depending on the parameters of interaction, these phases
may be separated by a region of gapless BP superfluid phase, Figure~\ref{picture5}.

Conditions to have BP phase are as follows.
At $p_{A}=p_{B}$ there is standard BCS, which is a stable fully gapped solution.
By adjusting the chemical potentials so as to increase the Fermi surface $p_{B}$,
we stress the system and low the pressurerelative to the normal phase.
Eventually, either before or after a transition to a BP state,
the pressure becomes negative and there is a first order phase transition
to the normal phase. At the point just before transition: if $\Delta_{p_{B}}$
is sufficiently large, the state is fully gapped (BCS) and no BP state
will occur; if $\Delta_{p_{B}}$ is small, then it will not appreciably
affect the dispersions and one finds a gapless Fermi surface coexisting
with the superfluid phase. As long as $\Delta_{p}$ falls off sufficiently quickly,
one can choose large ratio $m_{B}/m_{A}\gg 1$ so that the transition will occur
with $\Delta_{p_{B}}$ small enough to support the BP phase.
States shown at Figure~\ref{picture5} have $m_{B}/m_{A}=10$.
For a wider in q-space interaction, larger mass ratio is needed.

\section{Conclusion}

We considered a Fermi system with weak attractive interaction between species A and B, where
the BCS state forms at equal number densities, $n_{A}=n_{B}$.
What is the ground state of this system when $n_{A}\neq n_{B}$?
There is a range of parameters, where a breached pair phase exist. Breached
pair state is a homogeneous phase where superfluid and normal components
coexist. This state is stable and can be found provided there is a momentum
structure of interaction and large enough mass ratio of two species.

\section*{Acknowledgments}
This work is supported in part by funds provided by the U.S. Department
of Energy (D.O.E.) under cooperative research agreement DF-FC02-94ER40818.


\begin{thebibliography}{0}

\bibitem{Nature} For a review, see Nature 416, 205 (2002).

\bibitem{Review} For a review, see K Rajagopal and F. Wilczek,
hep-ph/0011333.

\bibitem{LiuWilczek} W. V. Liu and F. Wilczek, {\it Phys. Rev. Lett.}
{\bf 90}, 047002 (2003), cond-mat/0208052.

\bibitem{Sarma} G. Sarma, {\it Phys. Chem. Solid} {\bf 24}, 1029 (1963);
A. A. Abrikosov, ``Foundations of the theory of metalls''.

\bibitem{AbrikosovGorkovDzyaloshinski} A. A. Abrikosov, L. P. Gorkov,
and I. E. Dzyaloshinski, ``Methods of quantum field theory in statistical physics'',
Dover Publications, Inc., New York.

\bibitem{LiuWilczekZoller} W. V. Liu, F. Wilczek, and P. Zoller (2004), cond-mat/0404478.

\bibitem{AlfordKouvarisRajagopal} M. Alford, C. Kouvaris, and K. Rajagopal (2003),
hep-ph/0311286.

\bibitem{GubankovaLiuWilczek} E. Gubankova, W. V. Liu, and F. Wilczek,
{\it Phys. Rev. Lett.} {\bf 91}, 032001 (2003), hep-ph/0304016.

\bibitem{ShovkovyHuang} I. Shovkovy and M Huang, {\it Phys. Lett.} {\bf B564},
205 (2003), hep-ph/0302142.

\bibitem{BedaqueCaldasRupak} P. F. Bedaque, H. Caldas, and G Rupak,
{\it Phys. Rev. Lett.} {\bf 91}, 247002 (2003), cond-mat/0306694.

\bibitem{ForbesGubankovaLiuWilczek} M. M. Forbes, E. Gubankova, W. V. Liu, and
F. Wilczek, hep-ph/0405059.



\end{thebibliography}
\end{document}